\documentclass[aps, prl, reprint,twocolumn, citeautoscript, superscriptaddress, showkeys, showpacs]{revtex4-1}   %
\usepackage{xr}
\usepackage[fleqn,tbtags]{amsmath}
\usepackage{amsfonts, amssymb, amsxtra, textcomp}
\usepackage[]{graphicx}
\usepackage{grffile}
\usepackage{bbm}
\usepackage{bm}
\usepackage{color}
\usepackage{graphicx}
\usepackage[colorlinks=true,allcolors=blue,citecolor=blue,linkcolor=blue]{hyperref}

\allowdisplaybreaks
\usepackage{pdfpages}

\def\bmx{\begin{pmatrix}}
\def\emx{\end{pmatrix}}

\renewcommand{\approx}{\simeq}

\begin{document}

\title{Percolation \emph{via} combined electrostatic and chemical doping in complex oxide films}

\author{Peter P. Orth}
\affiliation{School of Physics and Astronomy, University of Minnesota, Minneapolis,
Minnesota 55455, USA}

\author{Rafael M. Fernandes}
\affiliation{School of Physics and Astronomy, University of Minnesota, Minneapolis,
Minnesota 55455, USA}

\author{Jeff Walter}
\affiliation{Department of Chemical Engineering and Materials Science, University
of Minnesota, Minneapolis, MN 55455, USA}

\author{C. Leighton}
\affiliation{Department of Chemical Engineering and Materials Science, University
of Minnesota, Minneapolis, MN 55455, USA}

\author{B. I. Shklovskii}
\affiliation{School of Physics and Astronomy, University of Minnesota, Minneapolis,
Minnesota 55455, USA}
\affiliation{Fine Theoretical Physics Institute, University of Minnesota, Minneapolis,
MN 55455, USA}

\date{\today}
\begin{abstract}
Stimulated by experimental advances in electrolyte gating methods, we investigate theoretically percolation in thin films of inhomogenous complex oxides, such as La$_{1-x}$Sr$_{x}$CoO$_{3}$ (LSCO), induced by a combination of bulk chemical and surface electrostatic doping. Using numerical and analytical methods, we identify two mechanisms that describe how bulk dopants reduce the amount of electrostatic surface charge required to reach percolation: (i) bulk-assisted surface percolation, and (ii) surface-assisted bulk percolation. We show that the critical surface charge strongly depends on the film thickness when the film is close to the chemical percolation threshold. In particular, thin films can be driven across the percolation transition by modest surface charge densities \emph{via} surface-assisted bulk percolation. If percolation is associated with the onset of ferromagnetism, as in LSCO, we further demonstrate that the presence of critical magnetic clusters extending from the film surface into the bulk results in considerable volume enhancement of the saturation magnetization, with pronounced experimental consequences. These results should significantly guide experimental work seeking to verify gate-induced percolation transitions in such materials.
\end{abstract}

\maketitle
\emph{Introduction.--} The rapidly growing field of complex oxide
heterostructures provides many opportunities for the observation of
new physical phenomena, with promising applications in future electronic
devices~\cite{doi:10.1146/annurev-conmatphys-062910-140445,NgaiWalkerAhn-AnnuRevMaterRes-2014,doi:10.1146/annurev-matsci-070813-113437}.
Examples include strain engineering to control structural and electronic
ground states~\cite{doi:10.1146/annurev-conmatphys-062910-140445,NgaiWalkerAhn-AnnuRevMaterRes-2014,doi:10.1146/annurev-matsci-070813-113437,Haeni-Nature-2004},
realization of novel two-dimensional (2D) electron gases at oxide
interfaces~\cite{doi:10.1146/annurev-matsci-070813-113437,OhtomoHwang-LAOSTO-Nature-2004,doi:10.1146/annurev-matsci-070813-113552},
and the observation of interfacial magnetic~\cite{doi:10.1146/annurev-conmatphys-062910-140445,NgaiWalkerAhn-AnnuRevMaterRes-2014,doi:10.1146/annurev-matsci-070813-113437,doi:10.1146/annurev-matsci-070813-113447}
and superconducting states~\cite{doi:10.1146/annurev-conmatphys-062910-140445,NgaiWalkerAhn-AnnuRevMaterRes-2014,doi:10.1146/annurev-matsci-070813-113437}.
Due to the lower charge carrier densities in these materials ($n\simeq10^{21}\text{cm}^{-3}$)
compared to conventional metals ($n\simeq10^{23}\text{cm}^{-3}$),
surface electrostatic or electrochemical control of these novel properties
\emph{via} the electric field effect also becomes an exciting possibility~\cite{NgaiWalkerAhn-AnnuRevMaterRes-2014,Ahn2006,AhnTrisconeMannhard-Nature-2003,Goldman-AnnuRevMaterRes-2014}.

Stimulated by the above situation, high-$\kappa$ dielectrics, ferroelectric
gating, and electrolyte gating (primarily with ionic liquids and gels)
have been successfully employed to electrostatically induce and control
large charge densities in these materials~\cite{NgaiWalkerAhn-AnnuRevMaterRes-2014,Ahn2006,AhnTrisconeMannhard-Nature-2003,Goldman-AnnuRevMaterRes-2014}.
Particularly prominent recent progress has been made with ionic liquid
and gel gating, the surface carrier densities achieved routinely exceeding
$s\simeq 10^{14} \text{cm}^{-2}$, corresponding to modulation of significant
fractions of an electron (or hole) per unit cell~\cite{NgaiWalkerAhn-AnnuRevMaterRes-2014,Ahn2006,AhnTrisconeMannhard-Nature-2003,Goldman-AnnuRevMaterRes-2014}.
This has, for example, enabled reversible external electrical control
of oxide electronic phase transitions from insulating to metallic~\cite{:/content/aip/journal/apl/97/14/10.1063/1.3496458,ADMA:ADMA201003241,Nakano-Nature-2012,Jeong1402-Science-2012},
to a superconducting state~\cite{UenoKawasaki-NatMat-2008,Bollinger-Nature-2011,PhysRevLett.107.027001},
or from paramagnetic to magnetically-ordered phases~\cite{ADMA:ADMA200900278,Yamada1065-Science-2012}.
Nevertheless, attainment of sufficient charge density to induce the
phase transitions of interest remains a challenge in many cases, due
to the need for $s\simeq10^{15}\text{cm}^{-2}$. In such cases one obvious
strategy is to employ a combination of chemical and electrostatic
doping, bringing the material close to some electronic/magnetic phase
boundary by chemical substitution, then using surface electrostatic
tuning of the carrier density to reversibly traverse the critical
point.

\begin{figure}[b!]
\centering \includegraphics[width=0.8\linewidth]{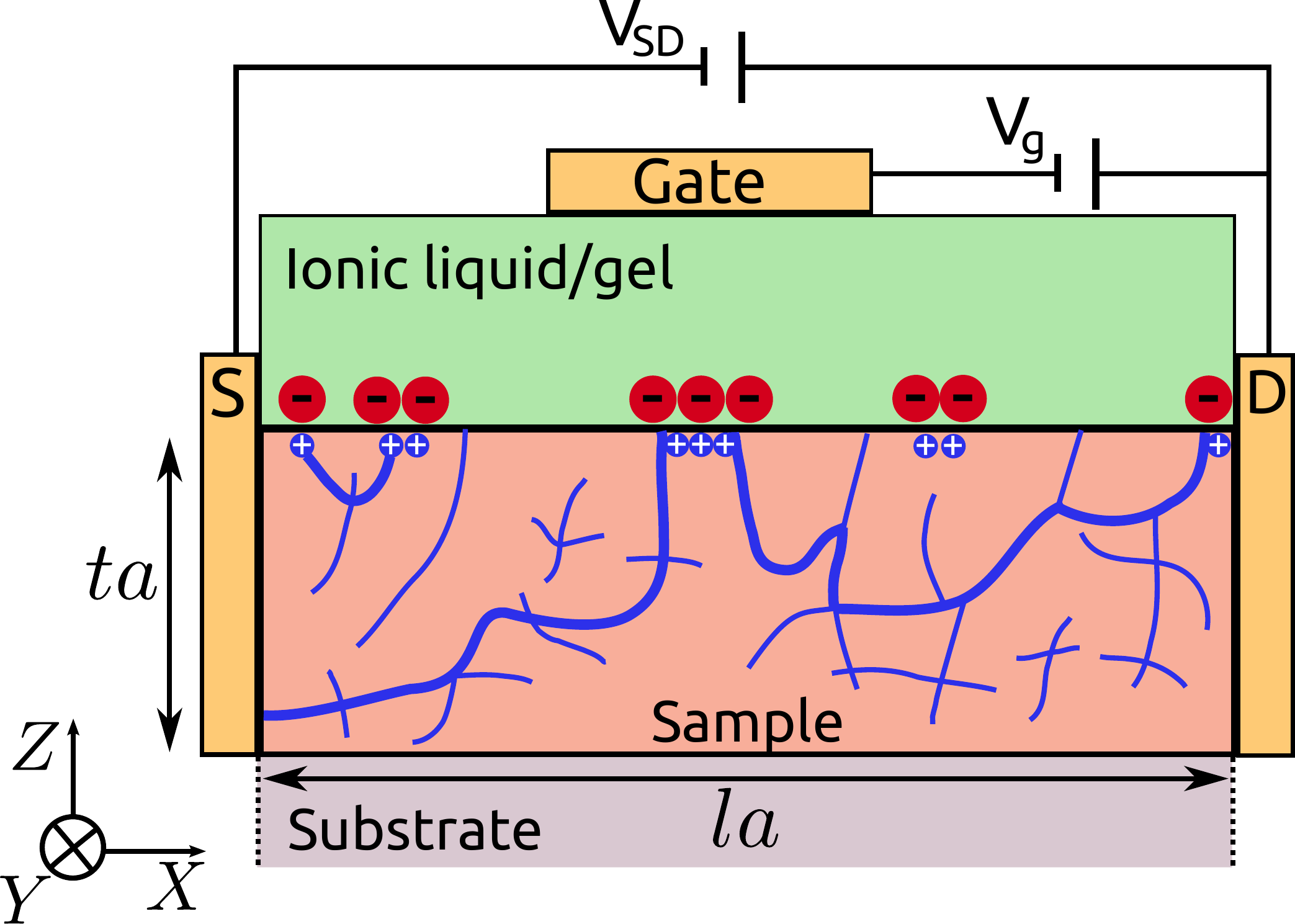}
\protect\caption{Schematic setup showing a thin film sample (red) of thickness $ta$
and area $la\times la$, where $a$ is the lattice constant, with
large finite clusters (blue) due to bulk doping. The ionic liquid or gel
(light green) on top of the sample induces a number of holes (blue
spheres) at the top surface -- proportional to the applied gate voltage
$V_{g}$. Red spheres denote anions in the ionic liquid/gel that move
towards the surface due to the applied voltage. For bulk doping close
to percolation $x_{c}-x\ll1$ (surface-assisted bulk percolation),
electrostatically induced holes connect finite bulk clusters at the
surface resulting in a conducting path (highlighted in the figure)
between source (S) and drain (D) electrodes. This leads to a current
driven by an applied source-drain voltage $V_{SD}$. The highlighted
upper left cluster shows bulk bridges connecting two surface clusters,
which is the dominant effect of bulk dopants for $x\ll x_{c}$ (bulk-assisted
surface percolation). }
\label{fig:1} 
\end{figure}

The work presented here focuses on exactly such combined electrostatic surface and bulk chemical doping. In particular, we investigate electronic/magnetic percolation transitions induced by a combination of chemical and electrostatic doping.
This is an important situation in complex oxide materials
due to the widespread observation of electronic and magnetic inhomogeneity
(as in manganites~\cite{Dagotto20011-Physics-Reports}, cuprates~\cite{Pasupathy196-Science-2008},
and cobaltites~\cite{PhysRevB.67.174408,0295-5075-87-2-27006} for
example), where many transitions, such as from insulator to metal
or from short- to long-range magnetism, are percolative in nature.
While our analysis and results are general, and could apply to percolation
transitions in various materials, in this paper we are motivated by physics
of the perovskite oxide cobaltite, La$_{1-x}$Sr$_{x}$CoO$_{3}$
(LSCO), which is well established to undergo a percolation transition
from insulator to metal at $x_{c}\approx0.18$~\cite{PhysRevB.67.174408,0295-5075-87-2-27006,aip/journal/apl/95/22/10.1063/1.3269192}.

In the parent compound LaCoO$_3$ ($x=0$), the Co$^{3+}$ ($3d^6$) ions adopt the $S=0$ spin state as $T \rightarrow 0$, and the material is a diamagnetic semiconductor. Substituting Sr$^{2+}$ for La$^{3+}$ induces holes, changing the formal valence state of a neighboring Co ion to $4+$, which is in a $S>0$ spin-state. The subsequent evolution from insulator to metal (due to hole transfer between nearest-neighbor Co$^{4+}$) and short- to long-range ferromagnetic correlations is caused by percolation of nanoscopic ferromagnetic hole-rich clusters~\cite{PhysRevB.67.174408, 0295-5075-87-2-27006, aip/journal/apl/95/22/10.1063/1.3269192}.
Very thin (few unit cell thick) films of LSCO are the natural target for field-effect gating experiments, as significant modulation of the charge carrier density is confined to a narrow layer close to the surface. The layer width is of the order of the electrostatic screening length, which is typically one or two unit cells~\cite{NgaiWalkerAhn-AnnuRevMaterRes-2014,Ahn2006,AhnTrisconeMannhard-Nature-2003,Goldman-AnnuRevMaterRes-2014} due to the large carrier densities ($n\simeq10^{21}\text{cm}^{-3}$) in significantly doped LSCO.

The theoretical study of percolation phenomena in correlated systems has a long history~\cite{SkhlovskiiEfros-ElectronicPropertiesOfDopedSemiconductors-Book,StaufferAharony-Percolation-Book,Vojta_Schmalian05,Haas05,Bray_Ali06,Sandvik06,Hoyos07,Altman08,Fernandes_Schmalian11}. However, the combination of bulk chemical and surface electrostatic doping defines an interesting and unusual percolation problem that is so far largely unexplored theoretically. The schematic setup with gate, source and drain electrodes is shown in Fig.~\ref{fig:1}, where the blue parts in the LSCO film denote hole-rich regions and
the top surface is affected by electrostatic gating. The total (top)
surface carrier density, 
\begin{align}
s=x+\Delta s\label{eq:5}
\end{align}
arises from doping both by chemical substitution of a fraction of lattice sites $x$ and electrostatic gating of a fraction of surface lattice sites $\Delta s$.

In this work we identify two different percolation phenomena: bulk-assisted surface percolation and surface-assisted bulk percolation, which are schematically depicted in Fig.~\ref{fig:1}. In the first case, where the system is initially far away from the (thickness-dependent) bulk percolation threshold $x_{c}(t)$, percolation on the surface is facilitated by diluted bulk dopants, which provide bridges that connect disjunct finite surface clusters. As a result, the amount of surface charge $\Delta s_{c}$ that must be induced electrostatically to reach percolation is insensitive to the film thickness. In the second case, where the bulk chemical doping level is close to the percolation threshold, $x_c(t) - x \ll 1$, we find that small $\Delta s$ helps to reach bulk percolation by connecting large finite bulk clusters on the surface. We show that the surface charge at percolation follows $s_c \propto t (x_c - x)$ for films of thickness $ta$. As a result, $\Delta s_c$ grows moderately with $(x_c-x)$ for thin films, but increases sharply for thicker films. In the particular case where the percolation transition is associated with ferromagnetic order, as in LSCO, the presence of clusters, which extend from the surface into the bulk, greatly enhance the surface saturation magnetization $M_{s}$. 

\emph{Numerical modeling of percolation.--} To derive our results, we consider the site percolation problem on the cubic lattice of size $la\times la\times ta$
along the $X$, $Y$ and $Z$ axes defined in Fig. \ref{fig:1}, where
$a$ is the lattice constant and $l,t$ are integers ($t\leq l$).
This geometry describes films of thickness $ta$ and surface
area $(la)^{2}$.
The percolation problem is solved using the numerical algorithm described
in Refs.~\onlinecite{PhysRevLett.85.4104,PhysRevE.64.016706}.
Starting from an empty lattice, a fraction $x$ of sites are first
randomly filled in the whole lattice to simulate bulk chemical doping.
We verify that the bulk doping percolation threshold on the
isotropic cubic lattice ($l=t$) lies at $x_{c}^{3D}=0.31$~\cite{PhysRevE.87.052107},
and increases for $t<l$, \emph{i.e.}, $x_{c}(t)>x_{c}(l)\equiv x_{c}^{3D}$~\cite{SkhlovskiiEfros-ElectronicPropertiesOfDopedSemiconductors-Book}.
To study the role of surface doping, we stop at a bulk doping level
$x<x_{c}(t)$ and subsequently add a fraction $\Delta s$ of sites
exclusively on the top surface layer to simulate electrostatic gating.
The total surface density of sites at the top surface is then given
by Eq.~\eqref{eq:5}. While electrostatically doping the system,
we continuously monitor whether a percolating path exists between
the two side surfaces at $X=0$ and $X=la$. We define the critical
total density of sites at the top surface that is required for percolation
between the side surfaces as $s_{c}$. The amount of charge density
that must be transferred \emph{via} electrostatic doping is then denoted
$\Delta s_{c}$. 

\begin{figure}[t!]
\centering 
\includegraphics[width=1\linewidth]{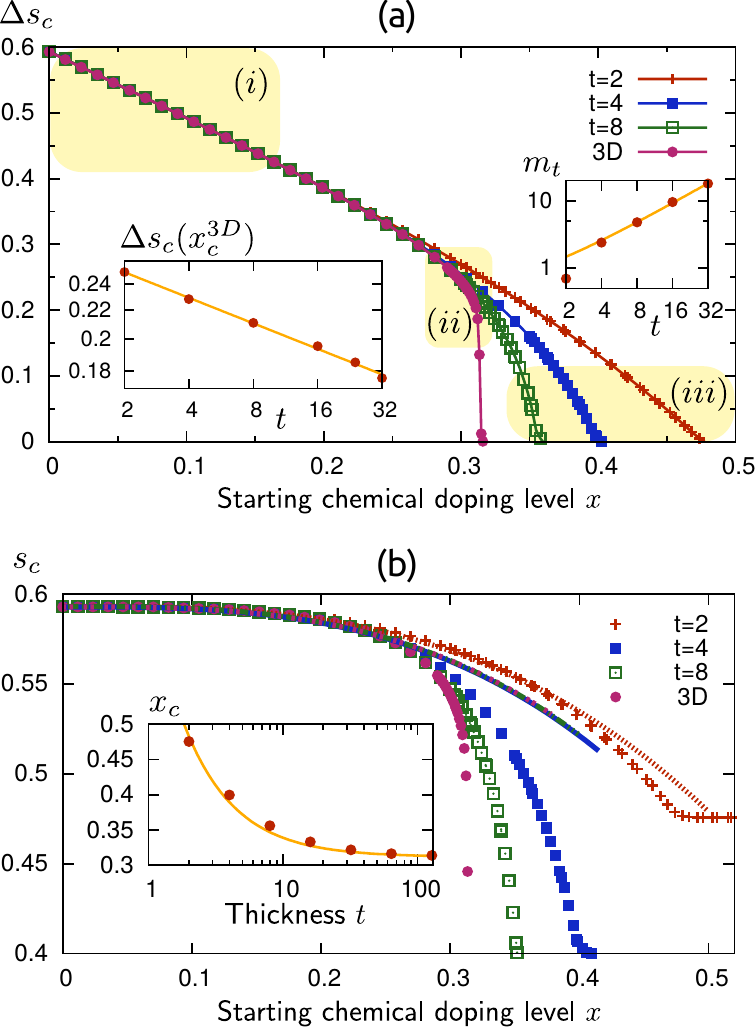} 
\caption{(a) Surface charge density $\Delta s_{c}$ that must be electrostatically induced to reach percolation, as a function of starting bulk chemical
doping level, $x$. Different curves correspond to different thicknesses $t$, as indicated, and are obtained from extrapolating results for system sizes $l\times l\times t$ with $l=32,64,128$ to $l^{-1}\rightarrow0$ and are averaged over at least $4.1\times10^{5}$ disorder realizations. The curve labelled ``3D'' is for $t=l$. The left inset shows that $\Delta s_{c}$ at the bulk percolation threshold $x_{c}^{3D}=0.31$ obeys Eq.~\eqref{eq:6} (yellow line) with $c_2=0.27$. The right inset shows the slope of $s_c-x_c = m_t (x_c - x)$ close to $x_c(t) - x \ll 1$ verifying Eq.~\eqref{eq:4} with $c_5 = 0.56$. Yellow rectangles mark the three regimes labelled (i), (ii), and (iii), addressed by our analytical theory. 
(b) Total surface charge at percolation, $s_{c}$, as a function of $x$. The lines are fits of the numerical results according to Eq.~\eqref{eq:2} with $b=0.91$ for $t=2$ and $b=1.12$ for $t=4,8, l$. The inset shows the thickness-dependent bulk percolation threshold $x_{c}(t)$ for purely chemical doping. The yellow line obeys Eq.~\eqref{eq:7} with $x_{c}^{3D}=0.312$, $\nu=0.88$~\cite{PhysRevE.87.052107} and $c_{3}=1.21$. }
\label{fig:2} 
\end{figure}

In Fig.~\ref{fig:2}(a), we show numerical results for $\Delta s_{c}$
as a function of the starting bulk chemical doping level $x$; panel (b) shows $s_c$ as a function of $x$. For
pure surface doping, $x=0$, we find the percolation threshold of
the 2D square lattice, $\Delta s_{c}(0)=0.59$~\cite{StaufferAharony-Percolation-Book}.
For small $x\ll x_{c}\left(t\right)$, the behavior of $\Delta s_{c}\left(x\right)$
depends only weakly on the film thickness $t$. In contrast, for $x_c(t) - x \ll 1$
the function $\Delta s_{c}\left(x\right)$ depends strongly on the
thickness $t$, displaying a sharp enhancement as $x$ decreases for
thick films but a much more gradual one for thin films. To
understand the numerical results, we next employ scaling theory arguments~\cite{SkhlovskiiEfros-ElectronicPropertiesOfDopedSemiconductors-Book}.

\emph{Analytical theory.-- }
To develop an analytical theory, we focus on three limits: (i) $x\ll x_{c}\left(t\right)$,
(ii) $x_{c}^{3D}-x\ll1$ and (iii) $x_{c}(t)-x\ll1$, which are indicated by yellow rectangles in Fig.~\ref{fig:2}(a). The first case
can be described as bulk-assisted surface percolation and the other two by surface-assisted bulk percolation.

(i) For $x\ll x_{c}\left(t\right)$, we have $s_{c}(0)-s_{c}(x)\ll1$:
the system is close to the 2D percolation threshold on the surface,
but far from percolation in the bulk. As a result, the typical size
of bulk clusters is rather small. These small bulk clusters can still
assist percolation at the surface by providing short bridges across
missing links between disconnected finite large surface clusters.
This situation is shown schematically in Fig.~\ref{fig:1}. 
Since the smallest possible bulk bridge consists of three sites below
the surface, at $x\ll1$ the main contribution of the bulk doping
arises from such bridges, yielding 
\begin{align}
s_{c}(x)=s_{c}(0)-bx^{3}\,.\label{eq:2}
\end{align}
As shown in Fig.~\ref{fig:2}(b), this equation, with weakly $t$-dependent
coefficient $b$, describes the numerical results well for $s_{c}(0)-s_{c}(x)\ll1$; for $t=2$ it is even applicable over almost the full range of doping levels up to $x_c$.

(ii) In the regime of small $x_{c}^{3D}-x\ll1$, the 3D bulk is close to the percolation threshold, but the surface concentration is far from the surface percolation threshold. Thus, while large critical finite clusters exist in the bulk, with a typical size of $\xi(x)\sim a(x_{c}^{3D}-x)^{-\nu}$ and correlation length exponent $\nu=0.88$~\cite{StaufferAharony-Percolation-Book,PhysRevE.87.052107}, the largest surface clusters remain small.

Let us first discuss the case of an infinite isotropic 3D system, before considering finite thickness films. If sites were randomly added in the bulk, an infinite cluster connecting $X=0$ and $X=la$, which looks like a network of links and nodes with typical separation $\xi(x)$, would occur after adding $N=N_{0}(x_{c}^{3D}-x)l^{3}$ sites, with $N_{0}\approx 2$. Because this infinite cluster provides percolation inside a layer of height $\xi(x)$ below the surface, the number of sites $\Delta N=N_{0}(x_{c}^{3D}-x)l^{2}\xi(x)/a$ we have added to this layer is sufficient to induce percolation along the layer. 
It is now plausible that instead of homogeneously doping the sliver
of volume $(la)^{2}\xi(x)$, we can reach percolation by adding all
these sites to the surface plane only. For a 3D system, this yields
a critical surface density of 
\begin{align}
s_{c}(x) & =x_{c}^{3D}+\frac{\Delta N}{l^{2}}=x_{c}^{3D}+c_{1}(x_{c}^{3D}-x)^{1-\nu}\,,\label{eq:3}
\end{align}
with a non-universal constant $c_{1}$ and $x_{c}^{3D}-x\ll1$. We
see that since $\nu<1$, connecting bulk clusters on the surface can
be done by very small surface addition $\Delta s$ at $x_{c}^{3D}-x\ll1$.
Of course, the scaling behavior in Eq.~\eqref{eq:3} only holds as
long as $(x_{c}^{3D}-x)^{1-\nu}\ll1$. Since $1-\nu=0.12\ll1$~\cite{StaufferAharony-Percolation-Book,PhysRevE.87.052107},
the validity of Eq.~\eqref{eq:3} is therefore limited to a tiny region of $x$ close to $x_{c}$.
This explains the sharp enhancement of $\Delta s_{c}\left(x\right)$ observed in the 3D numerical results shown in Fig.~\ref{fig:2}.

A finite thickness $t$ of the film introduces another length scale,
which cuts off the scaling behavior of Eq.~\eqref{eq:3} as soon
as the correlation length becomes larger than the film thickness.
We will now show that for bulk doping levels such that $\xi(x)\geq ta$, Eq.~\eqref{eq:3} is replaced by 
\begin{align}
s_{c}(x)=x_{c}^{3D}+c_{2}t^{1-1/\nu}\,,
\label{eq:6}
\end{align}
with non-universal constant $c_{2}$. We numerically verify this scaling behavior at $x = x_c^{3D}$ as shown in the (left) inset of Fig.~\ref{fig:2}(a).
To derive Eq.~\eqref{eq:6}, we first notice that the bulk percolation threshold $x_{c}\left(t\right)$ of a film of thickness $t$ is reached when an infinite bulk cluster with correlation length $\xi[x_{c}(t)]\leq t a$ appears. From this it follows that~\cite{SkhlovskiiEfros-ElectronicPropertiesOfDopedSemiconductors-Book}:
\begin{align}
x_{c}(t)=x_{c}^{3D}+c_{3}t^{-1/\nu}\,,\label{eq:7}
\end{align}
with non-universal constant $c_{3}=1.21$, which is in agreement with our numerical
results shown in the inset of Fig.~\ref{fig:2}(b).
Therefore, to achieve percolation at $x \approx x_c^{3D}$, a film with width $t$ must acquire
$\Delta N=c_{4}\bigl(x_{c}(t)-x_{c}^{3D}\bigr)tl^{2}=c_{2}l^{2}t^{1-1/\nu}$
filled sites, where $c_{4}$ is a non-universal constant. As above,
we assume that we can reach the percolation threshold by bringing
all these sites into the surface plane by electrostatic gating, yielding
Eq.~\eqref{eq:6}. Note that Eq.~\eqref{eq:6} crosses over to Eq.~\eqref{eq:3}
at $\xi(x)=t a$. 

(iii) We now investigate $s_{c}$ for $x_{c}(t)-x\ll1$. In this regime, it holds that $\xi(x)>ta$ since the correlation length at $x_{c}(t)$ fulfills $\xi[x_{c}(t)]=ta$. In this case, we find that $\Delta N=(x_{c}(t)-x)l^{2}t$ sites should be added to the system in order to reach percolation, such that the critical surface percolation
threshold obeys  
\begin{align}
s_{c}(x) & =x_{c}(t) +c_{5}t(x_{c}(t)-x)
\label{eq:4}
\end{align}
with non-universal constant $c_{5}$. 
We demonstrate in the (right) inset of Fig.~\ref{fig:2}(a) that our numerical results follow this scaling relation of the slope $m_t = c_5 t$ with $c_5 = 0.56$. 
Note that the scaling breaks down for the thinnest system, $t=2$, which is instead described by Eq.~\eqref{eq:2} over the full range of bulk doping levels $x$ (see Fig.~\ref{fig:2}(b)).

The key insight from the combined numerical and analytical results is that bulk chemical doping largely reduces the amount of electrostatic surface charge $\Delta s_c$ required to reach percolation (compared to the 2D value) in a region of initial chemical doping levels $x_{c}^{3D} < x < x_c(t)$. In this regime, the critical surface charge $s_c$ scales with the thickness according to Eq.~\eqref{eq:4} and therefore grows quickly for thicker films. The underlying physical phenomenon is that less surface charge must be transferred by electrostatic gating if percolation is induced by connecting finite large bulk clusters on the surface rather than creating a percolating path that is confined to the surface alone. The width of this region $x_c - x_c^{3D} \propto t^{-1/\nu}$ rapidly narrows for thicker films. For smaller $x$ the dominant effect of the bulk dopants is to act as short bridges between disconnected surface clusters. This reduces the number of surface sites that must be filled to reach percolation only slightly compared to the 2D case, as described by Eq.~\eqref{eq:2}. 

\begin{figure}[t]
\centering \includegraphics[width=\linewidth]{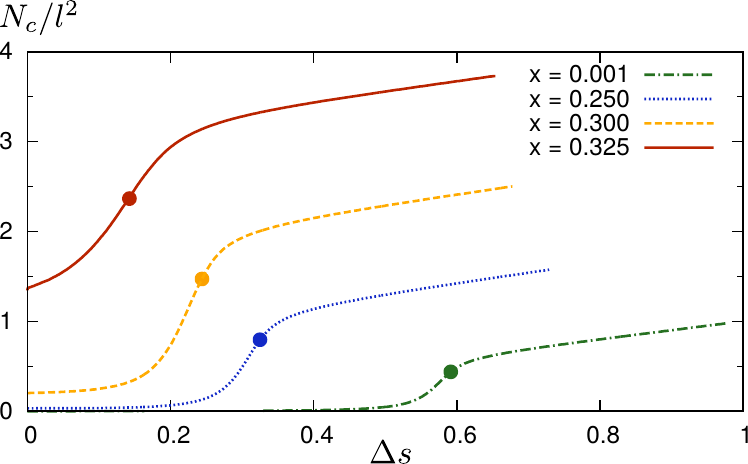}
\caption{Surface density of the largest cluster in the system, $N_{c}/l^{2}$,
as a function of electrostatic doping $\Delta s$ in a film of thickness
$t=16$. Dots indicate percolation thresholds. After crossing the
percolation threshold the largest cluster density is proportional
to the surface saturation magnetization $M_s$. The plot shows the
large (volume) enhancement of $M_s$, which occurs due to 
extension of the infinite cluster and its dead ends deep into the
bulk (see Fig. \ref{fig:1}).}
\label{fig:3} 
\end{figure}

\emph{Enhanced surface magnetization.--} If the percolation transition
is associated with ferromagnetic ordering, as for LSCO,
the extension of the percolating cluster from the surface into the
bulk leads to a dramatic volume enhancement of the surface saturation
magnetization $M_s$ in the case of surface-assisted bulk percolation (cases (ii) and (iii)).
To capture this phenomenon, in Fig.~\ref{fig:3} we show the size (\emph{i.e.} number of sites), of the largest cluster $N_{c}$ (per surface
area $l^{2}$) as a function of electrostatic doping $\Delta s$.
Beyond the percolation threshold $\Delta s>\Delta s_{c}(x)$ this
cluster percolates and its size is proportional to the surface saturation
magnetization $M_s\propto N_{c}/l^{2}$. Different curves correspond to different starting chemical doping
levels $0.003\leq x/x_{c}(t)\leq0.976$ and the film thickness is $t = 16$. 

For small doping levels, we observe
regular surface percolation at $\Delta s_{c}=0.59$ (the percolation
threshold is indicated by the dot). The percolated path is almost
entirely confined to the top surface layer and the magnetization enhancement
is absent: $N_{c}/l^{2}\lesssim1$. On the other hand, if the system
is initially doped closer to the (bulk) percolation threshold $x_{c}$,
the percolating cluster extends significantly into the bulk and we
observe $N_{c}/l^{2}>1$ for the other three realistic doping levels
$x$ we consider. As the (fractal) dimension of this cluster exceeds
$d=2$, we find that $N_{c}/l^{2}$ becomes as large as $4$ for a film of thickness $t=16$ (note that a fully magnetized film corresponds
to $N_{c}/l^{2}=t$). This shows that although bulk doping does not
assist greatly in \emph{reaching} percolation, it does ultimately
generate a much larger saturation magnetization in the sample, because
of the inclusion of preformed clusters of spin polarized sites (see
also Fig.~\ref{fig:1}). In addition, we further predict an unusual
depth profile of magnetization $M_s(z)$ as a function of distance
$z$ from the surface, which can be directly experimentally measured,
for instance using polarized neutron reflectometry. It could also
be indirectly inferred using perhaps x-ray magnetic circular dichroism
(XMCD) or the magneto-optical Kerr effect (MOKE).

\emph{Conclusions.--} Motivated by existing and ongoing experiments
on complex oxide thin films, we have studied a new percolation problem,
where bulk chemical doping is combined with electrostatic doping of
the surface. We have derived new analytical formulae describing universal
scaling behavior of the electrostatic percolation threshold and explored
the full crossover from bulk to surface percolation numerically. Experimental
predictions that follow from our analysis are that: (i) the critical
surface charge density at percolation $s_{c}$ depends only weakly
on the starting bulk doping level $x$, except in proximity to the bulk percolation transition $x_c^{3D} < x < x_c(t)$ . 
The crossover from surface-assisted to bulk-assisted percolation occurs
more abruptly for thicker films. Given limitations of ionic liquid/gel
or ferroelectric gating, experimental validations of gate-induced
percolation may thus rely in most cases on chemically doping close
to the percolation threshold. (ii) Once percolation is reached, the
saturation magnetization $M_s$ is largely enhanced due to the presence
of critical clusters extending deep into the bulk. 
 (iii) The existence of ferromagnetic bulk clusters will also be reflected
in the dependence of the magnetization $M_s(z)$ on the distance
$z$ from the surface. Our work thus shows that ``bulk'' magnetic
properties can be controlled using ``surface'' electrostatic gating.
We note that while the percolation threshold $x_{c}$ is a non-universal
quantity dependent on microscopic details such as the geometry of
the lattice, the scaling behavior of $s_{c}(x)$ that we
derive is \emph{universal}. Our results thus apply to LSCO and other
experimental systems even though the percolation threshold in this
material is not that of a simple cubic lattice $x^{3D}_{c} \approx 0.31$,
but rather $x^{3D}_{c,\text{LSCO}} \approx 0.18$. 
Finally, we note that while our analysis has focused
on effects of electrostatic gating, our conclusions also apply to
the case of electrochemical doping describing, for example, the transfer
of oxygen vacancies into the surface of a sample. 
\begin{acknowledgments}
We gratefully acknowledge useful discussions with K. Reich. This work
was supported primarily by the National Science Foundation through
the University of Minnesota MRSEC under Award Number DMR-1420013.
We acknowledge the Minnesota Supercomputing Institute (MSI) at the
University of Minnesota for providing resources that contributed to
the research results of this work. 
\end{acknowledgments}

\bibliographystyle{apsrev4-1}

\begin{thebibliography}{36}%
\makeatletter
\providecommand \@ifxundefined [1]{%
 \@ifx{#1\undefined}
}%
\providecommand \@ifnum [1]{%
 \ifnum #1\expandafter \@firstoftwo
 \else \expandafter \@secondoftwo
 \fi
}%
\providecommand \@ifx [1]{%
 \ifx #1\expandafter \@firstoftwo
 \else \expandafter \@secondoftwo
 \fi
}%
\providecommand \natexlab [1]{#1}%
\providecommand \enquote  [1]{``#1''}%
\providecommand \bibnamefont  [1]{#1}%
\providecommand \bibfnamefont [1]{#1}%
\providecommand \citenamefont [1]{#1}%
\providecommand \href@noop [0]{\@secondoftwo}%
\providecommand \href [0]{\begingroup \@sanitize@url \@href}%
\providecommand \@href[1]{\@@startlink{#1}\@@href}%
\providecommand \@@href[1]{\endgroup#1\@@endlink}%
\providecommand \@sanitize@url [0]{\catcode `\\12\catcode `\$12\catcode
  `\&12\catcode `\#12\catcode `\^12\catcode `\_12\catcode `\%12\relax}%
\providecommand \@@startlink[1]{}%
\providecommand \@@endlink[0]{}%
\providecommand \url  [0]{\begingroup\@sanitize@url \@url }%
\providecommand \@url [1]{\endgroup\@href {#1}{\urlprefix }}%
\providecommand \urlprefix  [0]{URL }%
\providecommand \Eprint [0]{\href }%
\providecommand \doibase [0]{http://dx.doi.org/}%
\providecommand \selectlanguage [0]{\@gobble}%
\providecommand \bibinfo  [0]{\@secondoftwo}%
\providecommand \bibfield  [0]{\@secondoftwo}%
\providecommand \translation [1]{[#1]}%
\providecommand \BibitemOpen [0]{}%
\providecommand \bibitemStop [0]{}%
\providecommand \bibitemNoStop [0]{.\EOS\space}%
\providecommand \EOS [0]{\spacefactor3000\relax}%
\providecommand \BibitemShut  [1]{\csname bibitem#1\endcsname}%
\let\auto@bib@innerbib\@empty
\bibitem [{\citenamefont {Zubko}\ \emph {et~al.}(2011)\citenamefont {Zubko},
  \citenamefont {Gariglio}, \citenamefont {Gabay}, \citenamefont {Ghosez},\
  and\ \citenamefont
  {Triscone}}]{doi:10.1146/annurev-conmatphys-062910-140445}%
  \BibitemOpen
  \bibfield  {author} {\bibinfo {author} {\bibfnamefont {P.}~\bibnamefont
  {Zubko}}, \bibinfo {author} {\bibfnamefont {S.}~\bibnamefont {Gariglio}},
  \bibinfo {author} {\bibfnamefont {M.}~\bibnamefont {Gabay}}, \bibinfo
  {author} {\bibfnamefont {P.}~\bibnamefont {Ghosez}}, \ and\ \bibinfo {author}
  {\bibfnamefont {J.-M.}\ \bibnamefont {Triscone}},\ }\href {\doibase
  10.1146/annurev-conmatphys-062910-140445} {\bibfield  {journal} {\bibinfo
  {journal} {Annu. Rev. Cond. Mat. Phys.}\ }\textbf {\bibinfo {volume} {2}},\
  \bibinfo {pages} {141} (\bibinfo {year} {2011})}\BibitemShut {NoStop}%
\bibitem [{\citenamefont {Ngai}\ \emph {et~al.}(2014)\citenamefont {Ngai},
  \citenamefont {Walker},\ and\ \citenamefont
  {Ahn}}]{NgaiWalkerAhn-AnnuRevMaterRes-2014}%
  \BibitemOpen
  \bibfield  {author} {\bibinfo {author} {\bibfnamefont {J.}~\bibnamefont
  {Ngai}}, \bibinfo {author} {\bibfnamefont {F.}~\bibnamefont {Walker}}, \ and\
  \bibinfo {author} {\bibfnamefont {C.}~\bibnamefont {Ahn}},\ }\href {\doibase
  10.1146/annurev-matsci-070813-113248} {\bibfield  {journal} {\bibinfo
  {journal} {Annu. Rev. Mater. Res.}\ }\textbf {\bibinfo {volume} {44}},\
  \bibinfo {pages} {1} (\bibinfo {year} {2014})}\BibitemShut {NoStop}%
\bibitem [{\citenamefont {Sulpizio}\ \emph {et~al.}(2014)\citenamefont
  {Sulpizio}, \citenamefont {Ilani}, \citenamefont {Irvin},\ and\ \citenamefont
  {Levy}}]{doi:10.1146/annurev-matsci-070813-113437}%
  \BibitemOpen
  \bibfield  {author} {\bibinfo {author} {\bibfnamefont {J.~A.}\ \bibnamefont
  {Sulpizio}}, \bibinfo {author} {\bibfnamefont {S.}~\bibnamefont {Ilani}},
  \bibinfo {author} {\bibfnamefont {P.}~\bibnamefont {Irvin}}, \ and\ \bibinfo
  {author} {\bibfnamefont {J.}~\bibnamefont {Levy}},\ }\href {\doibase
  10.1146/annurev-matsci-070813-113437} {\bibfield  {journal} {\bibinfo
  {journal} {Annu. Rev. Mater. Res.}\ }\textbf {\bibinfo {volume} {44}},\
  \bibinfo {pages} {117} (\bibinfo {year} {2014})}\BibitemShut {NoStop}%
\bibitem [{\citenamefont {Haeni}\ \emph {et~al.}(2004)\citenamefont {Haeni},
  \citenamefont {Irvin}, \citenamefont {Chang}, \citenamefont {Uecker},
  \citenamefont {Reiche}, \citenamefont {Li}, \citenamefont {Choudhury},
  \citenamefont {Tian}, \citenamefont {Hawley}, \citenamefont {Craigo},
  \citenamefont {Tagantsev}, \citenamefont {Pan}, \citenamefont {Streiffer},
  \citenamefont {Chen}, \citenamefont {Kirchoefer}, \citenamefont {Levy},\ and\
  \citenamefont {Schlom}}]{Haeni-Nature-2004}%
  \BibitemOpen
  \bibfield  {author} {\bibinfo {author} {\bibfnamefont {J.~H.}\ \bibnamefont
  {Haeni}}, \bibinfo {author} {\bibfnamefont {P.}~\bibnamefont {Irvin}},
  \bibinfo {author} {\bibfnamefont {W.}~\bibnamefont {Chang}}, \bibinfo
  {author} {\bibfnamefont {R.}~\bibnamefont {Uecker}}, \bibinfo {author}
  {\bibfnamefont {P.}~\bibnamefont {Reiche}}, \bibinfo {author} {\bibfnamefont
  {Y.~L.}\ \bibnamefont {Li}}, \bibinfo {author} {\bibfnamefont
  {S.}~\bibnamefont {Choudhury}}, \bibinfo {author} {\bibfnamefont
  {W.}~\bibnamefont {Tian}}, \bibinfo {author} {\bibfnamefont {M.~E.}\
  \bibnamefont {Hawley}}, \bibinfo {author} {\bibfnamefont {B.}~\bibnamefont
  {Craigo}}, \bibinfo {author} {\bibfnamefont {A.~K.}\ \bibnamefont
  {Tagantsev}}, \bibinfo {author} {\bibfnamefont {X.~Q.}\ \bibnamefont {Pan}},
  \bibinfo {author} {\bibfnamefont {S.~K.}\ \bibnamefont {Streiffer}}, \bibinfo
  {author} {\bibfnamefont {L.~Q.}\ \bibnamefont {Chen}}, \bibinfo {author}
  {\bibfnamefont {S.~W.}\ \bibnamefont {Kirchoefer}}, \bibinfo {author}
  {\bibfnamefont {J.}~\bibnamefont {Levy}}, \ and\ \bibinfo {author}
  {\bibfnamefont {D.~G.}\ \bibnamefont {Schlom}},\ }\href
  {http://dx.doi.org/10.1038/nature02773} {\bibfield  {journal} {\bibinfo
  {journal} {Nature}\ }\textbf {\bibinfo {volume} {430}},\ \bibinfo {pages}
  {758} (\bibinfo {year} {2004})}\BibitemShut {NoStop}%
\bibitem [{\citenamefont {Ohtomo}\ and\ \citenamefont
  {Hwang}(2004)}]{OhtomoHwang-LAOSTO-Nature-2004}%
  \BibitemOpen
  \bibfield  {author} {\bibinfo {author} {\bibfnamefont {A.}~\bibnamefont
  {Ohtomo}}\ and\ \bibinfo {author} {\bibfnamefont {H.}~\bibnamefont {Hwang}},\
  }\href@noop {} {\bibfield  {journal} {\bibinfo  {journal} {Nature}\ }\textbf
  {\bibinfo {volume} {427}},\ \bibinfo {pages} {423} (\bibinfo {year}
  {2004})}\BibitemShut {NoStop}%
\bibitem [{\citenamefont {Stemmer}\ and\ \citenamefont
  {Allen}(2014)}]{doi:10.1146/annurev-matsci-070813-113552}%
  \BibitemOpen
  \bibfield  {author} {\bibinfo {author} {\bibfnamefont {S.}~\bibnamefont
  {Stemmer}}\ and\ \bibinfo {author} {\bibfnamefont {S.~J.}\ \bibnamefont
  {Allen}},\ }\href {\doibase 10.1146/annurev-matsci-070813-113552} {\bibfield
  {journal} {\bibinfo  {journal} {Annu. Rev. Mater. Res.}\ }\textbf {\bibinfo
  {volume} {44}},\ \bibinfo {pages} {151} (\bibinfo {year} {2014})}\BibitemShut
  {NoStop}%
\bibitem [{\citenamefont {Bhattacharya}\ and\ \citenamefont
  {May}(2014)}]{doi:10.1146/annurev-matsci-070813-113447}%
  \BibitemOpen
  \bibfield  {author} {\bibinfo {author} {\bibfnamefont {A.}~\bibnamefont
  {Bhattacharya}}\ and\ \bibinfo {author} {\bibfnamefont {S.~J.}\ \bibnamefont
  {May}},\ }\href {\doibase 10.1146/annurev-matsci-070813-113447} {\bibfield
  {journal} {\bibinfo  {journal} {Annu. Rev. Mater. Res.}\ }\textbf {\bibinfo
  {volume} {44}},\ \bibinfo {pages} {65} (\bibinfo {year} {2014})}\BibitemShut
  {NoStop}%
\bibitem [{\citenamefont {Ahn}\ \emph {et~al.}(2006)\citenamefont {Ahn},
  \citenamefont {Bhattacharya}, \citenamefont {Di~Ventra}, \citenamefont
  {Eckstein}, \citenamefont {Frisbie}, \citenamefont {Gershenson},
  \citenamefont {Goldman}, \citenamefont {Inoue}, \citenamefont {Mannhart},
  \citenamefont {Millis}, \citenamefont {Morpurgo}, \citenamefont {Natelson},\
  and\ \citenamefont {Triscone}}]{Ahn2006}%
  \BibitemOpen
  \bibfield  {author} {\bibinfo {author} {\bibfnamefont {C.~H.}\ \bibnamefont
  {Ahn}}, \bibinfo {author} {\bibfnamefont {A.}~\bibnamefont {Bhattacharya}},
  \bibinfo {author} {\bibfnamefont {M.}~\bibnamefont {Di~Ventra}}, \bibinfo
  {author} {\bibfnamefont {J.~N.}\ \bibnamefont {Eckstein}}, \bibinfo {author}
  {\bibfnamefont {C.~D.}\ \bibnamefont {Frisbie}}, \bibinfo {author}
  {\bibfnamefont {M.~E.}\ \bibnamefont {Gershenson}}, \bibinfo {author}
  {\bibfnamefont {A.~M.}\ \bibnamefont {Goldman}}, \bibinfo {author}
  {\bibfnamefont {I.~H.}\ \bibnamefont {Inoue}}, \bibinfo {author}
  {\bibfnamefont {J.}~\bibnamefont {Mannhart}}, \bibinfo {author}
  {\bibfnamefont {A.~J.}\ \bibnamefont {Millis}}, \bibinfo {author}
  {\bibfnamefont {A.~F.}\ \bibnamefont {Morpurgo}}, \bibinfo {author}
  {\bibfnamefont {D.}~\bibnamefont {Natelson}}, \ and\ \bibinfo {author}
  {\bibfnamefont {J.-M.}\ \bibnamefont {Triscone}},\ }\href {\doibase
  10.1103/RevModPhys.78.1185} {\bibfield  {journal} {\bibinfo  {journal} {Rev.
  Mod. Phys.}\ }\textbf {\bibinfo {volume} {78}},\ \bibinfo {pages} {1185}
  (\bibinfo {year} {2006})}\BibitemShut {NoStop}%
\bibitem [{\citenamefont {Ahn}\ \emph {et~al.}(2003)\citenamefont {Ahn},
  \citenamefont {Triscone},\ and\ \citenamefont
  {Mannhart}}]{AhnTrisconeMannhard-Nature-2003}%
  \BibitemOpen
  \bibfield  {author} {\bibinfo {author} {\bibfnamefont {C.~H.}\ \bibnamefont
  {Ahn}}, \bibinfo {author} {\bibfnamefont {J.-M.}\ \bibnamefont {Triscone}}, \
  and\ \bibinfo {author} {\bibfnamefont {J.}~\bibnamefont {Mannhart}},\
  }\href@noop {} {\bibfield  {journal} {\bibinfo  {journal} {Nature (London)}\
  }\textbf {\bibinfo {volume} {424}},\ \bibinfo {pages} {1015} (\bibinfo {year}
  {2003})}\BibitemShut {NoStop}%
\bibitem [{\citenamefont {Goldman}(2014)}]{Goldman-AnnuRevMaterRes-2014}%
  \BibitemOpen
  \bibfield  {author} {\bibinfo {author} {\bibfnamefont {A.~M.}\ \bibnamefont
  {Goldman}},\ }\href@noop {} {\bibfield  {journal} {\bibinfo  {journal} {Annu.
  Rev. Mater. Res.}\ }\textbf {\bibinfo {volume} {44}},\ \bibinfo {pages} {45}
  (\bibinfo {year} {2014})}\BibitemShut {NoStop}%
\bibitem [{\citenamefont {Asanuma}\ \emph {et~al.}(2010)\citenamefont
  {Asanuma}, \citenamefont {Xiang}, \citenamefont {Yamada}, \citenamefont
  {Sato}, \citenamefont {Inoue}, \citenamefont {Akoh}, \citenamefont {Sawa},
  \citenamefont {Ueno}, \citenamefont {Shimotani}, \citenamefont {Yuan},
  \citenamefont {Kawasaki},\ and\ \citenamefont
  {Iwasa}}]{:/content/aip/journal/apl/97/14/10.1063/1.3496458}%
  \BibitemOpen
  \bibfield  {author} {\bibinfo {author} {\bibfnamefont {S.}~\bibnamefont
  {Asanuma}}, \bibinfo {author} {\bibfnamefont {P.-H.}\ \bibnamefont {Xiang}},
  \bibinfo {author} {\bibfnamefont {H.}~\bibnamefont {Yamada}}, \bibinfo
  {author} {\bibfnamefont {H.}~\bibnamefont {Sato}}, \bibinfo {author}
  {\bibfnamefont {I.~H.}\ \bibnamefont {Inoue}}, \bibinfo {author}
  {\bibfnamefont {H.}~\bibnamefont {Akoh}}, \bibinfo {author} {\bibfnamefont
  {A.}~\bibnamefont {Sawa}}, \bibinfo {author} {\bibfnamefont {K.}~\bibnamefont
  {Ueno}}, \bibinfo {author} {\bibfnamefont {H.}~\bibnamefont {Shimotani}},
  \bibinfo {author} {\bibfnamefont {H.}~\bibnamefont {Yuan}}, \bibinfo {author}
  {\bibfnamefont {M.}~\bibnamefont {Kawasaki}}, \ and\ \bibinfo {author}
  {\bibfnamefont {Y.}~\bibnamefont {Iwasa}},\ }\href@noop {} {\bibfield
  {journal} {\bibinfo  {journal} {Appl. Phys. Lett.}\ }\textbf {\bibinfo
  {volume} {97}},\ \bibinfo {eid} {142110} (\bibinfo {year}
  {2010})}\BibitemShut {NoStop}%
\bibitem [{\citenamefont {Scherwitzl}\ \emph {et~al.}(2010)\citenamefont
  {Scherwitzl}, \citenamefont {Zubko}, \citenamefont {Lezama}, \citenamefont
  {Ono}, \citenamefont {Morpurgo}, \citenamefont {Catalan},\ and\ \citenamefont
  {Triscone}}]{ADMA:ADMA201003241}%
  \BibitemOpen
  \bibfield  {author} {\bibinfo {author} {\bibfnamefont {R.}~\bibnamefont
  {Scherwitzl}}, \bibinfo {author} {\bibfnamefont {P.}~\bibnamefont {Zubko}},
  \bibinfo {author} {\bibfnamefont {I.~G.}\ \bibnamefont {Lezama}}, \bibinfo
  {author} {\bibfnamefont {S.}~\bibnamefont {Ono}}, \bibinfo {author}
  {\bibfnamefont {A.~F.}\ \bibnamefont {Morpurgo}}, \bibinfo {author}
  {\bibfnamefont {G.}~\bibnamefont {Catalan}}, \ and\ \bibinfo {author}
  {\bibfnamefont {J.-M.}\ \bibnamefont {Triscone}},\ }\href {\doibase
  10.1002/adma.201003241} {\bibfield  {journal} {\bibinfo  {journal} {Adv.
  Mater.}\ }\textbf {\bibinfo {volume} {22}},\ \bibinfo {pages} {5517}
  (\bibinfo {year} {2010})}\BibitemShut {NoStop}%
\bibitem [{\citenamefont {Nakano}\ \emph {et~al.}(2012)\citenamefont {Nakano},
  \citenamefont {Shibuya}, \citenamefont {Okuyama}, \citenamefont {Hatano},
  \citenamefont {Ono}, \citenamefont {Kawasaki}, \citenamefont {Iwasa},\ and\
  \citenamefont {Tokura}}]{Nakano-Nature-2012}%
  \BibitemOpen
  \bibfield  {author} {\bibinfo {author} {\bibfnamefont {M.}~\bibnamefont
  {Nakano}}, \bibinfo {author} {\bibfnamefont {K.}~\bibnamefont {Shibuya}},
  \bibinfo {author} {\bibfnamefont {D.}~\bibnamefont {Okuyama}}, \bibinfo
  {author} {\bibfnamefont {T.}~\bibnamefont {Hatano}}, \bibinfo {author}
  {\bibfnamefont {S.}~\bibnamefont {Ono}}, \bibinfo {author} {\bibfnamefont
  {M.}~\bibnamefont {Kawasaki}}, \bibinfo {author} {\bibfnamefont
  {Y.}~\bibnamefont {Iwasa}}, \ and\ \bibinfo {author} {\bibfnamefont
  {Y.}~\bibnamefont {Tokura}},\ }\href {http://dx.doi.org/10.1038/nature11296}
  {\bibfield  {journal} {\bibinfo  {journal} {Nature}\ }\textbf {\bibinfo
  {volume} {487}},\ \bibinfo {pages} {459} (\bibinfo {year}
  {2012})}\BibitemShut {NoStop}%
\bibitem [{\citenamefont {Jeong}\ \emph {et~al.}(2013)\citenamefont {Jeong},
  \citenamefont {Aetukuri}, \citenamefont {Graf}, \citenamefont {Schladt},
  \citenamefont {Samant},\ and\ \citenamefont
  {Parkin}}]{Jeong1402-Science-2012}%
  \BibitemOpen
  \bibfield  {author} {\bibinfo {author} {\bibfnamefont {J.}~\bibnamefont
  {Jeong}}, \bibinfo {author} {\bibfnamefont {N.}~\bibnamefont {Aetukuri}},
  \bibinfo {author} {\bibfnamefont {T.}~\bibnamefont {Graf}}, \bibinfo {author}
  {\bibfnamefont {T.~D.}\ \bibnamefont {Schladt}}, \bibinfo {author}
  {\bibfnamefont {M.~G.}\ \bibnamefont {Samant}}, \ and\ \bibinfo {author}
  {\bibfnamefont {S.~S.~P.}\ \bibnamefont {Parkin}},\ }\href {\doibase
  10.1126/science.1230512} {\bibfield  {journal} {\bibinfo  {journal}
  {Science}\ }\textbf {\bibinfo {volume} {339}},\ \bibinfo {pages} {1402}
  (\bibinfo {year} {2013})}\BibitemShut {NoStop}%
\bibitem [{\citenamefont {Ueno}\ \emph {et~al.}(2008)\citenamefont {Ueno},
  \citenamefont {Nakamura}, \citenamefont {Shimotani}, \citenamefont {Ohtomo},
  \citenamefont {Kimura}, \citenamefont {Nojima}, \citenamefont {Aoki},
  \citenamefont {Iwasa},\ and\ \citenamefont
  {Kawasaki}}]{UenoKawasaki-NatMat-2008}%
  \BibitemOpen
  \bibfield  {author} {\bibinfo {author} {\bibfnamefont {K.}~\bibnamefont
  {Ueno}}, \bibinfo {author} {\bibfnamefont {S.}~\bibnamefont {Nakamura}},
  \bibinfo {author} {\bibfnamefont {H.}~\bibnamefont {Shimotani}}, \bibinfo
  {author} {\bibfnamefont {A.}~\bibnamefont {Ohtomo}}, \bibinfo {author}
  {\bibfnamefont {N.}~\bibnamefont {Kimura}}, \bibinfo {author} {\bibfnamefont
  {T.}~\bibnamefont {Nojima}}, \bibinfo {author} {\bibfnamefont
  {H.}~\bibnamefont {Aoki}}, \bibinfo {author} {\bibfnamefont {Y.}~\bibnamefont
  {Iwasa}}, \ and\ \bibinfo {author} {\bibfnamefont {M.}~\bibnamefont
  {Kawasaki}},\ }\href@noop {} {\bibfield  {journal} {\bibinfo  {journal} {Nat.
  Mat.}\ }\textbf {\bibinfo {volume} {7}},\ \bibinfo {pages} {855} (\bibinfo
  {year} {2008})}\BibitemShut {NoStop}%
\bibitem [{\citenamefont {Bollinger}\ \emph {et~al.}(2011)\citenamefont
  {Bollinger}, \citenamefont {Dubuis}, \citenamefont {Yoon}, \citenamefont
  {Pavuna}, \citenamefont {Misewich},\ and\ \citenamefont
  {Bozovic}}]{Bollinger-Nature-2011}%
  \BibitemOpen
  \bibfield  {author} {\bibinfo {author} {\bibfnamefont {A.~T.}\ \bibnamefont
  {Bollinger}}, \bibinfo {author} {\bibfnamefont {G.}~\bibnamefont {Dubuis}},
  \bibinfo {author} {\bibfnamefont {J.}~\bibnamefont {Yoon}}, \bibinfo {author}
  {\bibfnamefont {D.}~\bibnamefont {Pavuna}}, \bibinfo {author} {\bibfnamefont
  {J.}~\bibnamefont {Misewich}}, \ and\ \bibinfo {author} {\bibfnamefont
  {I.}~\bibnamefont {Bozovic}},\ }\href {http://dx.doi.org/10.1038/nature09998}
  {\bibfield  {journal} {\bibinfo  {journal} {Nature}\ }\textbf {\bibinfo
  {volume} {472}},\ \bibinfo {pages} {458} (\bibinfo {year}
  {2011})}\BibitemShut {NoStop}%
\bibitem [{\citenamefont {Leng}\ \emph {et~al.}(2011)\citenamefont {Leng},
  \citenamefont {Garcia-Barriocanal}, \citenamefont {Bose}, \citenamefont
  {Lee},\ and\ \citenamefont {Goldman}}]{PhysRevLett.107.027001}%
  \BibitemOpen
  \bibfield  {author} {\bibinfo {author} {\bibfnamefont {X.}~\bibnamefont
  {Leng}}, \bibinfo {author} {\bibfnamefont {J.}~\bibnamefont
  {Garcia-Barriocanal}}, \bibinfo {author} {\bibfnamefont {S.}~\bibnamefont
  {Bose}}, \bibinfo {author} {\bibfnamefont {Y.}~\bibnamefont {Lee}}, \ and\
  \bibinfo {author} {\bibfnamefont {A.~M.}\ \bibnamefont {Goldman}},\ }\href
  {\doibase 10.1103/PhysRevLett.107.027001} {\bibfield  {journal} {\bibinfo
  {journal} {Phys. Rev. Lett.}\ }\textbf {\bibinfo {volume} {107}},\ \bibinfo
  {pages} {027001} (\bibinfo {year} {2011})}\BibitemShut {NoStop}%
\bibitem [{\citenamefont {Molegraaf}\ \emph {et~al.}(2009)\citenamefont
  {Molegraaf}, \citenamefont {Hoffman}, \citenamefont {Vaz}, \citenamefont
  {Gariglio}, \citenamefont {van~der Marel}, \citenamefont {Ahn},\ and\
  \citenamefont {Triscone}}]{ADMA:ADMA200900278}%
  \BibitemOpen
  \bibfield  {author} {\bibinfo {author} {\bibfnamefont {H.~J.~A.}\
  \bibnamefont {Molegraaf}}, \bibinfo {author} {\bibfnamefont {J.}~\bibnamefont
  {Hoffman}}, \bibinfo {author} {\bibfnamefont {C.~A.~F.}\ \bibnamefont {Vaz}},
  \bibinfo {author} {\bibfnamefont {S.}~\bibnamefont {Gariglio}}, \bibinfo
  {author} {\bibfnamefont {D.}~\bibnamefont {van~der Marel}}, \bibinfo {author}
  {\bibfnamefont {C.~H.}\ \bibnamefont {Ahn}}, \ and\ \bibinfo {author}
  {\bibfnamefont {J.-M.}\ \bibnamefont {Triscone}},\ }\href {\doibase
  10.1002/adma.200900278} {\bibfield  {journal} {\bibinfo  {journal} {Adv.
  Mater.}\ }\textbf {\bibinfo {volume} {21}},\ \bibinfo {pages} {3470}
  (\bibinfo {year} {2009})}\BibitemShut {NoStop}%
\bibitem [{\citenamefont {Yamada}\ \emph {et~al.}(2011)\citenamefont {Yamada},
  \citenamefont {Ueno}, \citenamefont {Fukumura}, \citenamefont {Yuan},
  \citenamefont {Shimotani}, \citenamefont {Iwasa}, \citenamefont {Gu},
  \citenamefont {Tsukimoto}, \citenamefont {Ikuhara},\ and\ \citenamefont
  {Kawasaki}}]{Yamada1065-Science-2012}%
  \BibitemOpen
  \bibfield  {author} {\bibinfo {author} {\bibfnamefont {Y.}~\bibnamefont
  {Yamada}}, \bibinfo {author} {\bibfnamefont {K.}~\bibnamefont {Ueno}},
  \bibinfo {author} {\bibfnamefont {T.}~\bibnamefont {Fukumura}}, \bibinfo
  {author} {\bibfnamefont {H.~T.}\ \bibnamefont {Yuan}}, \bibinfo {author}
  {\bibfnamefont {H.}~\bibnamefont {Shimotani}}, \bibinfo {author}
  {\bibfnamefont {Y.}~\bibnamefont {Iwasa}}, \bibinfo {author} {\bibfnamefont
  {L.}~\bibnamefont {Gu}}, \bibinfo {author} {\bibfnamefont {S.}~\bibnamefont
  {Tsukimoto}}, \bibinfo {author} {\bibfnamefont {Y.}~\bibnamefont {Ikuhara}},
  \ and\ \bibinfo {author} {\bibfnamefont {M.}~\bibnamefont {Kawasaki}},\
  }\href {\doibase 10.1126/science.1202152} {\bibfield  {journal} {\bibinfo
  {journal} {Science}\ }\textbf {\bibinfo {volume} {332}},\ \bibinfo {pages}
  {1065} (\bibinfo {year} {2011})}\BibitemShut {NoStop}%
\bibitem [{\citenamefont {Dagotto}\ \emph {et~al.}(2001)\citenamefont
  {Dagotto}, \citenamefont {Hotta},\ and\ \citenamefont
  {Moreo}}]{Dagotto20011-Physics-Reports}%
  \BibitemOpen
  \bibfield  {author} {\bibinfo {author} {\bibfnamefont {E.}~\bibnamefont
  {Dagotto}}, \bibinfo {author} {\bibfnamefont {T.}~\bibnamefont {Hotta}}, \
  and\ \bibinfo {author} {\bibfnamefont {A.}~\bibnamefont {Moreo}},\ }\href
  {\doibase http://dx.doi.org/10.1016/S0370-1573(00)00121-6} {\bibfield
  {journal} {\bibinfo  {journal} {Phys. Rep.}\ }\textbf {\bibinfo {volume}
  {344}},\ \bibinfo {pages} {1 } (\bibinfo {year} {2001})}\BibitemShut
  {NoStop}%
\bibitem [{\citenamefont {Pasupathy}\ \emph {et~al.}(2008)\citenamefont
  {Pasupathy}, \citenamefont {Pushp}, \citenamefont {Gomes}, \citenamefont
  {Parker}, \citenamefont {Wen}, \citenamefont {Xu}, \citenamefont {Gu},
  \citenamefont {Ono}, \citenamefont {Ando},\ and\ \citenamefont
  {Yazdani}}]{Pasupathy196-Science-2008}%
  \BibitemOpen
  \bibfield  {author} {\bibinfo {author} {\bibfnamefont {A.~N.}\ \bibnamefont
  {Pasupathy}}, \bibinfo {author} {\bibfnamefont {A.}~\bibnamefont {Pushp}},
  \bibinfo {author} {\bibfnamefont {K.~K.}\ \bibnamefont {Gomes}}, \bibinfo
  {author} {\bibfnamefont {C.~V.}\ \bibnamefont {Parker}}, \bibinfo {author}
  {\bibfnamefont {J.}~\bibnamefont {Wen}}, \bibinfo {author} {\bibfnamefont
  {Z.}~\bibnamefont {Xu}}, \bibinfo {author} {\bibfnamefont {G.}~\bibnamefont
  {Gu}}, \bibinfo {author} {\bibfnamefont {S.}~\bibnamefont {Ono}}, \bibinfo
  {author} {\bibfnamefont {Y.}~\bibnamefont {Ando}}, \ and\ \bibinfo {author}
  {\bibfnamefont {A.}~\bibnamefont {Yazdani}},\ }\href {\doibase
  10.1126/science.1154700} {\bibfield  {journal} {\bibinfo  {journal}
  {Science}\ }\textbf {\bibinfo {volume} {320}},\ \bibinfo {pages} {196}
  (\bibinfo {year} {2008})}\BibitemShut {NoStop}%
\bibitem [{\citenamefont {Wu}\ and\ \citenamefont
  {Leighton}(2003)}]{PhysRevB.67.174408}%
  \BibitemOpen
  \bibfield  {author} {\bibinfo {author} {\bibfnamefont {J.}~\bibnamefont
  {Wu}}\ and\ \bibinfo {author} {\bibfnamefont {C.}~\bibnamefont {Leighton}},\
  }\href {\doibase 10.1103/PhysRevB.67.174408} {\bibfield  {journal} {\bibinfo
  {journal} {Phys. Rev. B}\ }\textbf {\bibinfo {volume} {67}},\ \bibinfo
  {pages} {174408} (\bibinfo {year} {2003})}\BibitemShut {NoStop}%
\bibitem [{\citenamefont {He}\ \emph {et~al.}(2009{\natexlab{a}})\citenamefont
  {He}, \citenamefont {El-Khatib}, \citenamefont {Wu}, \citenamefont {Lynn},
  \citenamefont {Zheng}, \citenamefont {Mitchell},\ and\ \citenamefont
  {Leighton}}]{0295-5075-87-2-27006}%
  \BibitemOpen
  \bibfield  {author} {\bibinfo {author} {\bibfnamefont {C.}~\bibnamefont
  {He}}, \bibinfo {author} {\bibfnamefont {S.}~\bibnamefont {El-Khatib}},
  \bibinfo {author} {\bibfnamefont {J.}~\bibnamefont {Wu}}, \bibinfo {author}
  {\bibfnamefont {J.~W.}\ \bibnamefont {Lynn}}, \bibinfo {author}
  {\bibfnamefont {H.}~\bibnamefont {Zheng}}, \bibinfo {author} {\bibfnamefont
  {J.~F.}\ \bibnamefont {Mitchell}}, \ and\ \bibinfo {author} {\bibfnamefont
  {C.}~\bibnamefont {Leighton}},\ }\href
  {http://stacks.iop.org/0295-5075/87/i=2/a=27006} {\bibfield  {journal}
  {\bibinfo  {journal} {EPL}\ }\textbf {\bibinfo {volume} {87}},\ \bibinfo
  {pages} {27006} (\bibinfo {year} {2009}{\natexlab{a}})}\BibitemShut {NoStop}%
\bibitem [{\citenamefont {He}\ \emph {et~al.}(2009{\natexlab{b}})\citenamefont
  {He}, \citenamefont {El-Khatib}, \citenamefont {Eisenberg}, \citenamefont
  {Manno}, \citenamefont {Lynn}, \citenamefont {Zheng}, \citenamefont
  {Mitchell},\ and\ \citenamefont
  {Leighton}}]{aip/journal/apl/95/22/10.1063/1.3269192}%
  \BibitemOpen
  \bibfield  {author} {\bibinfo {author} {\bibfnamefont {C.}~\bibnamefont
  {He}}, \bibinfo {author} {\bibfnamefont {S.}~\bibnamefont {El-Khatib}},
  \bibinfo {author} {\bibfnamefont {S.}~\bibnamefont {Eisenberg}}, \bibinfo
  {author} {\bibfnamefont {M.}~\bibnamefont {Manno}}, \bibinfo {author}
  {\bibfnamefont {J.~W.}\ \bibnamefont {Lynn}}, \bibinfo {author}
  {\bibfnamefont {H.}~\bibnamefont {Zheng}}, \bibinfo {author} {\bibfnamefont
  {J.~F.}\ \bibnamefont {Mitchell}}, \ and\ \bibinfo {author} {\bibfnamefont
  {C.}~\bibnamefont {Leighton}},\ }\href@noop {} {\bibfield  {journal}
  {\bibinfo  {journal} {Appl. Phys. Lett.}\ }\textbf {\bibinfo {volume} {95}},\
  \bibinfo {eid} {222511} (\bibinfo {year} {2009}{\natexlab{b}})}\BibitemShut
  {NoStop}%
\bibitem [{\citenamefont {Shklovskii}\ and\ \citenamefont
  {Efros}(1984)}]{SkhlovskiiEfros-ElectronicPropertiesOfDopedSemiconductors-Book}%
  \BibitemOpen
  \bibfield  {author} {\bibinfo {author} {\bibfnamefont {B.~I.}\ \bibnamefont
  {Shklovskii}}\ and\ \bibinfo {author} {\bibfnamefont {A.~L.}\ \bibnamefont
  {Efros}},\ }\href@noop {} {\emph {\bibinfo {title} {Electronic Properties of
  Doped Semiconductors}}},\ \bibinfo {series} {Springer Series in Solid-State
  Sciences}, Vol.~\bibinfo {volume} {45}\ (\bibinfo  {publisher} {Springer},\
  \bibinfo {address} {Heidelberg},\ \bibinfo {year} {1984})\BibitemShut
  {NoStop}%
\bibitem [{\citenamefont {Stauffer}\ and\ \citenamefont
  {Aharony}(1994)}]{StaufferAharony-Percolation-Book}%
  \BibitemOpen
  \bibfield  {author} {\bibinfo {author} {\bibfnamefont {D.}~\bibnamefont
  {Stauffer}}\ and\ \bibinfo {author} {\bibfnamefont {A.}~\bibnamefont
  {Aharony}},\ }\href@noop {} {\emph {\bibinfo {title} {Introduction to
  Percolation Theory}}},\ \bibinfo {edition} {2nd}\ ed.\ (\bibinfo  {publisher}
  {Taylor \& Francis},\ \bibinfo {address} {London (UK)},\ \bibinfo {year}
  {1994})\BibitemShut {NoStop}%
\bibitem [{\citenamefont {Vojta}\ and\ \citenamefont
  {Schmalian}(2005)}]{Vojta_Schmalian05}%
  \BibitemOpen
  \bibfield  {author} {\bibinfo {author} {\bibfnamefont {T.}~\bibnamefont
  {Vojta}}\ and\ \bibinfo {author} {\bibfnamefont {J.}~\bibnamefont
  {Schmalian}},\ }\href {\doibase 10.1103/PhysRevLett.95.237206} {\bibfield
  {journal} {\bibinfo  {journal} {Phys. Rev. Lett.}\ }\textbf {\bibinfo
  {volume} {95}},\ \bibinfo {pages} {237206} (\bibinfo {year}
  {2005})}\BibitemShut {NoStop}%
\bibitem [{\citenamefont {Yu}\ \emph {et~al.}(2005)\citenamefont {Yu},
  \citenamefont {Roscilde},\ and\ \citenamefont {Haas}}]{Haas05}%
  \BibitemOpen
  \bibfield  {author} {\bibinfo {author} {\bibfnamefont {R.}~\bibnamefont
  {Yu}}, \bibinfo {author} {\bibfnamefont {T.}~\bibnamefont {Roscilde}}, \ and\
  \bibinfo {author} {\bibfnamefont {S.}~\bibnamefont {Haas}},\ }\href {\doibase
  10.1103/PhysRevLett.94.197204} {\bibfield  {journal} {\bibinfo  {journal}
  {Phys. Rev. Lett.}\ }\textbf {\bibinfo {volume} {94}},\ \bibinfo {pages}
  {197204} (\bibinfo {year} {2005})}\BibitemShut {NoStop}%
\bibitem [{\citenamefont {Bray-Ali}\ \emph {et~al.}(2006)\citenamefont
  {Bray-Ali}, \citenamefont {Moore}, \citenamefont {Senthil},\ and\
  \citenamefont {Vishwanath}}]{Bray_Ali06}%
  \BibitemOpen
  \bibfield  {author} {\bibinfo {author} {\bibfnamefont {N.}~\bibnamefont
  {Bray-Ali}}, \bibinfo {author} {\bibfnamefont {J.~E.}\ \bibnamefont {Moore}},
  \bibinfo {author} {\bibfnamefont {T.}~\bibnamefont {Senthil}}, \ and\
  \bibinfo {author} {\bibfnamefont {A.}~\bibnamefont {Vishwanath}},\ }\href
  {\doibase 10.1103/PhysRevB.73.064417} {\bibfield  {journal} {\bibinfo
  {journal} {Phys. Rev. B}\ }\textbf {\bibinfo {volume} {73}},\ \bibinfo
  {pages} {064417} (\bibinfo {year} {2006})}\BibitemShut {NoStop}%
\bibitem [{\citenamefont {Wang}\ and\ \citenamefont
  {Sandvik}(2006)}]{Sandvik06}%
  \BibitemOpen
  \bibfield  {author} {\bibinfo {author} {\bibfnamefont {L.}~\bibnamefont
  {Wang}}\ and\ \bibinfo {author} {\bibfnamefont {A.~W.}\ \bibnamefont
  {Sandvik}},\ }\href {\doibase 10.1103/PhysRevLett.97.117204} {\bibfield
  {journal} {\bibinfo  {journal} {Phys. Rev. Lett.}\ }\textbf {\bibinfo
  {volume} {97}},\ \bibinfo {pages} {117204} (\bibinfo {year}
  {2006})}\BibitemShut {NoStop}%
\bibitem [{\citenamefont {Hoyos}\ \emph {et~al.}(2007)\citenamefont {Hoyos},
  \citenamefont {Kotabage},\ and\ \citenamefont {Vojta}}]{Hoyos07}%
  \BibitemOpen
  \bibfield  {author} {\bibinfo {author} {\bibfnamefont {J.~A.}\ \bibnamefont
  {Hoyos}}, \bibinfo {author} {\bibfnamefont {C.}~\bibnamefont {Kotabage}}, \
  and\ \bibinfo {author} {\bibfnamefont {T.}~\bibnamefont {Vojta}},\ }\href
  {\doibase 10.1103/PhysRevLett.99.230601} {\bibfield  {journal} {\bibinfo
  {journal} {Phys. Rev. Lett.}\ }\textbf {\bibinfo {volume} {99}},\ \bibinfo
  {pages} {230601} (\bibinfo {year} {2007})}\BibitemShut {NoStop}%
\bibitem [{\citenamefont {Altman}\ \emph {et~al.}(2008)\citenamefont {Altman},
  \citenamefont {Kafri}, \citenamefont {Polkovnikov},\ and\ \citenamefont
  {Refael}}]{Altman08}%
  \BibitemOpen
  \bibfield  {author} {\bibinfo {author} {\bibfnamefont {E.}~\bibnamefont
  {Altman}}, \bibinfo {author} {\bibfnamefont {Y.}~\bibnamefont {Kafri}},
  \bibinfo {author} {\bibfnamefont {A.}~\bibnamefont {Polkovnikov}}, \ and\
  \bibinfo {author} {\bibfnamefont {G.}~\bibnamefont {Refael}},\ }\href
  {\doibase 10.1103/PhysRevLett.100.170402} {\bibfield  {journal} {\bibinfo
  {journal} {Phys. Rev. Lett.}\ }\textbf {\bibinfo {volume} {100}},\ \bibinfo
  {pages} {170402} (\bibinfo {year} {2008})}\BibitemShut {NoStop}%
\bibitem [{\citenamefont {Fernandes}\ and\ \citenamefont
  {Schmalian}(2011)}]{Fernandes_Schmalian11}%
  \BibitemOpen
  \bibfield  {author} {\bibinfo {author} {\bibfnamefont {R.~M.}\ \bibnamefont
  {Fernandes}}\ and\ \bibinfo {author} {\bibfnamefont {J.}~\bibnamefont
  {Schmalian}},\ }\href {\doibase 10.1103/PhysRevLett.106.067004} {\bibfield
  {journal} {\bibinfo  {journal} {Phys. Rev. Lett.}\ }\textbf {\bibinfo
  {volume} {106}},\ \bibinfo {pages} {067004} (\bibinfo {year}
  {2011})}\BibitemShut {NoStop}%
\bibitem [{\citenamefont {Newman}\ and\ \citenamefont
  {Ziff}(2000)}]{PhysRevLett.85.4104}%
  \BibitemOpen
  \bibfield  {author} {\bibinfo {author} {\bibfnamefont {M.~E.~J.}\
  \bibnamefont {Newman}}\ and\ \bibinfo {author} {\bibfnamefont {R.~M.}\
  \bibnamefont {Ziff}},\ }\href {\doibase 10.1103/PhysRevLett.85.4104}
  {\bibfield  {journal} {\bibinfo  {journal} {Phys. Rev. Lett.}\ }\textbf
  {\bibinfo {volume} {85}},\ \bibinfo {pages} {4104} (\bibinfo {year}
  {2000})}\BibitemShut {NoStop}%
\bibitem [{\citenamefont {Newman}\ and\ \citenamefont
  {Ziff}(2001)}]{PhysRevE.64.016706}%
  \BibitemOpen
  \bibfield  {author} {\bibinfo {author} {\bibfnamefont {M.~E.~J.}\
  \bibnamefont {Newman}}\ and\ \bibinfo {author} {\bibfnamefont {R.~M.}\
  \bibnamefont {Ziff}},\ }\href {\doibase 10.1103/PhysRevE.64.016706}
  {\bibfield  {journal} {\bibinfo  {journal} {Phys. Rev. E}\ }\textbf {\bibinfo
  {volume} {64}},\ \bibinfo {pages} {016706} (\bibinfo {year}
  {2001})}\BibitemShut {NoStop}%
\bibitem [{\citenamefont {Wang}\ \emph {et~al.}(2013)\citenamefont {Wang},
  \citenamefont {Zhou}, \citenamefont {Zhang}, \citenamefont {Garoni},\ and\
  \citenamefont {Deng}}]{PhysRevE.87.052107}%
  \BibitemOpen
  \bibfield  {author} {\bibinfo {author} {\bibfnamefont {J.}~\bibnamefont
  {Wang}}, \bibinfo {author} {\bibfnamefont {Z.}~\bibnamefont {Zhou}}, \bibinfo
  {author} {\bibfnamefont {W.}~\bibnamefont {Zhang}}, \bibinfo {author}
  {\bibfnamefont {T.~M.}\ \bibnamefont {Garoni}}, \ and\ \bibinfo {author}
  {\bibfnamefont {Y.}~\bibnamefont {Deng}},\ }\href {\doibase
  10.1103/PhysRevE.87.052107} {\bibfield  {journal} {\bibinfo  {journal} {Phys.
  Rev. E}\ }\textbf {\bibinfo {volume} {87}},\ \bibinfo {pages} {052107}
  (\bibinfo {year} {2013})}\BibitemShut {NoStop}%
\end{thebibliography}

%

\end{document}